\documentclass[useAMS,usenatbib]{mn2e}
\usepackage{amsmath,fleqn,graphicx,amssymb}
\usepackage{multirow}
\def\bolth{\mbox{\boldmath$\theta$}}
\def\Gyr{\,\mathrm{Gyr}}
\def\kpc{\,\mathrm{kpc}}
\def\kms{\,\mathrm{km\,s}^{-1}}
\def\Gevdens{\,\mathrm{GeV\,cm}^{-3}}
\def\msun{\,{\rm M}_\odot}

\def\pc{\,\mathrm{pc}}
\def\rmd{\,\mathrm{d}}
\def\vsol{\mathbf{v_\odot}}

\title[Mass models of the Milky Way]
{Mass models of the Milky Way}

\author[P.~J.~McMillan]{
  Paul~J.~McMillan\thanks{E-mail: p.mcmillan1@physics.ox.ac.uk}\\
  Rudolf Peierls Centre for Theoretical Physics, 1 Keble Road,
  Oxford, OX1 3NP, UK
}

\begin{document}
\maketitle

\begin{abstract}
  We present a simple method for fitting parametrized mass models of
  the Milky Way to observational constraints. We take a Bayesian
  approach which allows us to take into account input from photometric
  and kinematic data, and expectations from theoretical modelling.
  This provides us with a best-fitting model, which is a suitable
  starting point for dynamical modelling. We also determine a
  probability density function on the properties of the model, which
  demonstrates that the mass distribution of the Galaxy remains very
  uncertain. For our choices of parametrization and constraints, we
  find disc scale lengths of $3.00\pm0.22\kpc$ and $3.29\pm0.56\kpc$
  for the thin and thick discs respectively; a Solar radius of 
  $8.29\pm0.16\kpc$ 
  and a circular speed at the Sun of $239\pm5\kms$; a total stellar mass of
  $6.43\pm0.63\times10^{10}\msun$; a virial mass of
  $1.26\pm0.24\times10^{12}\msun$ and a local dark matter density of
  $0.40\pm0.04\Gevdens$. We find some correlations between the 
  best-fitting parameters of our models (for example, between the disk scale 
  lengths and the Solar radius), which we discuss. The chosen 
  disc scale-heights are shown to have little effect on the key properties 
  of the model.
\end{abstract}


\section{Introduction}\label{sec:intro}
A great deal is still unknown about the distribution of mass in the
various components of the Milky Way. The major discoveries in Galactic
astronomy over the past decade have almost all been related to
components which comprise a small fraction of the total mass of the
Milky Way, most of which either are or were dwarf galaxies \citep[for
example, the many objects observed in the ``Field of
Streams'',][]{FieldofStreams_short}.  The structure of the dominant
components -- the disc(s) and cold dark matter (CDM) halo -- remains
rather uncertain.

An important element of understanding and constraining the structure of the 
major components of the Galaxy is creating Galaxy models which can be 
compared to observational data.
It is important to draw a distinction between three types of Galaxy
models: mass, kinematic and dynamical models. Mass models are the
simplest of these, and only attempt to describe the density
distribution of the various Galaxy components, and thus the Galactic
potential \citep*[e.g.][henceforth DB98]{KlZhSu02,WDJJB98:mass}. 
Kinematic models, such as those produced by \textsc{galaxia}
\citep{Galaxia}, specify the density and velocity distributions of the
luminous components of 
the Galaxy, but do not consider the question of whether these are consistent
with a steady state in any Galactic potential. Dynamical models 
\citep*[e.g.][]{WiPyDu08} describe
systems which are in a steady state in a given potential, because
their distribution functions depend only on the integrals of
motion. The Besan\c{c}on Galaxy model \citep{Roea03} is primarily a 
kinematic model with a dynamical element used to determine the vertical 
structure of the disc.

It is clear that moving beyond simple kinematic models to full dynamical 
ones is an essential step in fully understanding our Galaxy. The majority of 
the mass of the Galaxy is expected to lie in the CDM halo, which is
only observable through its gravitational effect on luminous components of 
the Galaxy, so purely kinematic models cannot provide any insight into its 
structure. 

The first step towards a dynamical model is to produce a mass model that 
is consistent with available constraints. 
An influential mass model was that of \cite{Sc56}, and, as observational data 
and understanding of galaxy structure improved, updated versions have been 
produced by other authors, notably \cite{CaOs81} and DB98. Our 
intention in this study is to follow these authors in producing a mass
model that is consistent with up-to-date observational data and 
theoretical understanding, and to provide a simple framework for producing
these models into which future data can be placed as they become available. 

The major difficulty in producing a model of this kind is drawing
together data from numerous different studies of the various
components which make up the Milky Way in a way that is as consistent
as possible. Such studies often make different underlying assumptions,
and sometimes come to seemingly mutually contradictory conclusions.
In principle the correct approach is to return to the raw data that
each study was based upon and to synthesise them into one
coherent picture.  Even if this is possible in practice, it is
certainly an immense undertaking, and one we do not attempt
here. Instead we follow the approach of previous authors by 
accepting various constraints on the parameters of our
models as stated by other studies, without returning to the raw data.
In addition we use well understood kinematic data
sets, and -- for the CDM halo, about which observational data is
limited -- we use our best current theoretical understanding.  Our
approach is best thought of as Bayesian with direct constraints on the
model parameters (from photometric data or theoretical insight) being
our Bayesian ``prior'', and kinematic data used to find the
likelihood.

In this paper we present a simple method for determining both a
best-fitting parametrized mass model of the Galaxy, and
the full probability density function of the parameters of the model. 
This paper is associated with a series of papers in which we construct dynamical
models to fit observational data 
\citep*[][McMillan et al. in preparation]{JJB10,JJBPJM11:dyn}.  
Eventually these dynamical
models will themselves be used to constrain the Galactic potential.  

Like DB98 we restrict ourselves to axisymmetric models. It
is clear that the Galaxy is not actually axisymmetric, especially the
inner Galaxy (see Section~\ref{sec:bulge}), but the disc is close to
axisymmetric \citep{Juea08_short}, and (as noted by DB98) axisymmetric
models successfully account for observations in the 21-cm line of
hydrogen at Galactic longitudes $l\gtrsim30$, corresponding to
$R\gtrsim4\kpc$.

To find the gravitational potential associated with a given mass model
we use the publicly available code \textsc{galpot}, which is described
by DB98 section 2.3.

In Section~\ref{sec:comp} we describe the different components that
make up the bulk of the mass of the Galaxy, and how our model
represents them. We also explain our Bayesian priors on the parameters
that describe our model. In Section~\ref{sec:data} we give the
kinematic data we use to constrain our model. All the constraints 
applied to the model are summarised in Table~\ref{tab:constr}.
In Section~\ref{sec:fit},
we describe the method used to fit the model, and in Section~\ref{sec:res} we
give our best-fitting model and the rest of our results. In 
Section~\ref{sec:compare} we compare our results to other data sets.

Unless otherwise stated, any measurement or derived quantity we use to
constrain our model is assumed to have Gaussian uncertainties.

\section{Components of the Milky Way}\label{sec:comp}
\subsection{The bulge}\label{sec:bulge}

The Galactic bulge has been shown in many studies to be a
near-prolate, triaxial rotating bar with its long axis in the plane of
the Galaxy \citep*[e.g.][]{JJBGeSp97}.  However, our interest in this
study is in producing an axisymmetric model, so we are forced to make
crude approximations in our modelling of this component. We must
therefore accept that it is unwise to compare our model to
measurements taken from the inner few $\kpc$ of the Galaxy, as it
can be expected to do a poor job of reproducing them.

Insight into the structure of the bulge can be gained from photometric
studies, however one must be careful when doing so as there can be a
major contribution to the stellar density in the inner few $\kpc$ from
the disc component. The model used for the disc in these studies will
therefore have a significant effect on the properties determined for
the bulge. This goes some way towards explaining why the mass of the
bulge as determined by \cite{PiRo04}, $2.4\pm0.6\times10^{10}\msun$,
using a photometric model with a ``hole'' in the centre of the disc,
is so much larger than that determined by studies using kinematic data
\citep*[e.g. DB98;][]{WiPyDu08,BiGe02} which do not include a
disc ``hole'' in their models.

Our density profile is based on the parametric model which
\cite{BiGe02} fit to dereddened L-band COBE/DIRBE data
\citep*{SpMaBl96}, and the mass-to-light ratio that
\citeauthor{BiGe02} determine from a comparison between gas dynamics
in models and those observed in the inner Galaxy. This model is found
on the assumption that the disc component has no central hole. This also
assumes that the mass-to-light ratio is spatially constant, which
allows us to convert a photometric model directly into a mass model
for this component.

The \citeauthor{BiGe02} model is not axisymmetric, so we make an
axisymmetrised approximation which has the density profile
\begin{equation} \label{eq:bulge}
  \rho_\mathrm{b}=\frac{\rho_{\mathrm{b},0}}{(1+r^\prime/r_0)^\alpha}\;
  \textrm{exp}\left[-\left(r^\prime/r_{\mathrm{cut}}\right)^2\right],
\end{equation}
where, in cylindrical coordinates,
\begin{equation}
  r^\prime = \sqrt{R^2 + (z/q)^2}
\end{equation}
with $\alpha=1.8$, $r_0=0.075\mathrm{kpc}$, $r_{\mathrm{cut}}=2.1\mathrm{kpc}$,
and axis ratio $q=0.5$. The \citeauthor{BiGe02} mass-to-light ratio
has a quoted uncertainty $\pm5\%$, but given the extent to which we
have altered this model it is prudent to recognize the need to
introduce further uncertainty in our model fitting. We therefore
assume that the bulge mass $M_\mathrm{b}=8.9\times 10^9\,\msun$, with
uncertainty $\pm10\%$. For this density profile, this corresponds to
scale density $\rho_{\mathrm{b},0}=9.93\times 10^{10}\,\msun\kpc^{-3}
\pm 10\%$.

\subsection{The disc}\label{sec:disc}
The Milky Way's disc is usually considered to have two major components:
a thin disc and a thick disc \citep[e.g.][]{GiRe83}. These are
generally modelled as exponential in the sense that
\begin{equation} \label{eq:disc}
  \rho_\mathrm{d}(R,z)=\frac{\Sigma_{\mathrm{d},0}}{2z_\mathrm{d}}\;\textrm{exp}\left(-\frac{\mid
      z\mid}{z_\mathrm{d}}-\frac{R}{R_\mathrm{d}}\right),
\end{equation}
with scale height $z_\mathrm{d}$, scale length $R_\mathrm{d}$ and central surface
density $\Sigma_{\mathrm{d},0}$. The total mass of a disc like this is
$M_\mathrm{d}=2\pi\Sigma_{\mathrm{d},0}R_\mathrm{d}^2$.

As mentioned in Section~\ref{sec:bulge}, some studies of the inner Galaxy 
prefer models with a central ``hole'' in the disc \citep[e.g.][]{PiRo04}. We 
do not consider such models here. The kinematic data we consider 
(Section~\ref{sec:data}) is all related to parts of the Galaxy which lie 
outside any central disc hole, so, in this study, a model with a central
hole should simply redistribute mass in the inner few $\kpc$ from the 
disc to the bulge 
(note that our prior on the bulge would have to be replaced, as it is taken 
from a study which used a model of the disc with no central hole).

The \cite{Juea08_short} analysis of data from the Sloan Digital Sky Survey
\citep[SDSS:][]{SDSS7_short} showed that the approximation to
exponential profiles is a sensible one for the Milky Way, and produced
estimates based on photometry for the scale lengths, scale heights and
relative densities of the two discs.

The scale-heights of the discs are not at all well constrained by the
kinematic data we use in this study, so initially we accept without question the
\cite{Juea08_short} best-fitting values, $z_{\mathrm{d},\mathrm{thin}}=300\pc$ and 
$z_{\mathrm{d},\mathrm{thick}}=900\pc$. In Section~\ref{sec:height} we explore the
(relatively small) effects of changing the assumed disc scale-heights. 

The \citeauthor{Juea08_short} scale-lengths for the thin and thick
discs are $2.6$ and $3.6\kpc$ respectively, with a quoted uncertainty of
$20\%$ in each case.  The local density normalisation $f_{\mathrm{d},\odot} =
\rho_{\mathrm{thick}}(R_\odot,z_\odot)/ \rho_{\mathrm{thin}}(R_\odot,z_\odot)$ is 
quoted
as $0.12$, with uncertainty $10\%$. We approximate these uncertainties
as Gaussian and uncorrelated, and take these as prior probability
distributions on $R_{\mathrm{d},\mathrm{thin}}$, $R_{\mathrm{d},\mathrm{thick}}$ 
and $f_{\mathrm{d},\odot}$.  Again, we
assume a constant mass-to-light ratio which allows us to convert these
photometric constraints directly into constraints on the mass density.

\subsection{The dark-matter halo} \label{sec:halo} For obvious
reasons, we cannot use photometric data to constrain 
the shape of the dark matter distribution, so
instead we look to cosmological simulations for insight.

In cosmological simulations that only include dark matter, halo
density profiles are well fit by a universal profile, known as the NFW
profile \citep*{NFW96}
\begin{equation} \label{eq:NFW} \rho_{\mathrm{h}} =
  \frac{\rho_{\mathrm{h},0}}{x\,(1+x)^2},
\end{equation}
where $x=r/r_{\mathrm{h}}$, with $r_\mathrm{h}$ the scale radius.  It is clear that the
baryons within CDM haloes will have an impact on the halo
profile, but the nature of this impact remains uncertain because it 
depends on the complex physics of baryons. 
Cosmological simulations suggest that the condensation of baryons to
the centre of galactic haloes will cause the density profile in the
inner parts of the halo to be steeper than $\rho\propto r^{-1}$, as it
is in the NFW case \citep{Duea10}. Observations of low surface
brightness galaxies, however, appear to indicate that they lie in 
dark-matter haloes with constant density cores. This difference between
observation and simulations (with or without baryonic physics) is
known as the ``cusp-core problem'' \citep[for an overview
see][]{dB10}.

Observations of low surface brightness galaxies and dwarf galaxies can
used to provide more reliable information about the inner dark matter
profile than galaxies like the Milky Way because they are dark matter
dominated right to their centre, so the mass contribution of the
baryons can be modelled with little uncertainty compared to the total
mass.  The Milky Way is not dark matter dominated and it is very hard
to determine slope of the dark matter density profile in the inner
Galaxy because the baryonic density is dominant in the inner Galaxy,
and still uncertain.

Haloes in dark-matter-only cosmological simulations tend to be
significantly prolate, but with a great deal of variation in axis
ratios \citep[e.g.][]{Alea06}. It is recognised that, again, baryonic
physics will play an important role -- condensation of baryons to the
centres of haloes is expected to make them rounder than dark-matter-only 
simulations would suggest \citep{Deea08}.  The shape of the Milky
Way's halo is still very much the subject of debate, with different
efforts to fit models of the Sagittarius dwarf's orbit favouring
conflicting halo shapes \citep*[see e.g.][]{LaMaJo09}.

In this study we do not attempt to address the question of the shape
of the Milky Way halo, or the impact of baryonic physics on the CDM
density profile.  We take the simple model of a spherically symmetric
NFW halo (eq.~\ref{eq:NFW}). As better understanding of the effect of
baryonic physics on CDM haloes or constraints from observations become
available, these simple assumptions will have to be revisited.

It is common to describe CDM haloes in terms of their virial
mass $M_{\mathrm{\mathrm{vir}}}$ -- which is the mass contained within the virial radius
$r_{\mathrm{\mathrm{vir}}}$ -- and the concentration $c=r_{-2}/r_{\mathrm{vir}}$, where $r_{-2}$
is the radius at which the logarithmic slope of the density profile
$\rmd \log\rho/\rmd\log r = -2$ (for an NFW profile,
$r_{-2}=r_\mathrm{h}$). The virial radius $r_{\mathrm{\mathrm{vir}}}$ is defined as the radius of
a sphere centred on the halo centre which has an average density of
$\Delta$ times the critical density, but the definition of $\Delta$
varies between authors. The definitions that are relevant to this
paper are $\Delta=200$ (when using this definition we will use the
notation $M_{v}$, $r_{v}$ and $c_{v}$ for the virial mass, radius, and
the concentration), and a value of $\Delta$ which varies with
redshift, derived from spherical top-hat collapse models
\citep[e.g.][]{GuGo72}, with $\Delta\approx94$ today (for which we use
the notation $M_{v^\prime}$, $r_{v^\prime}$, $c_{v^\prime}$).

\cite{Guea10} compared the distribution of halo virial masses found in
the Millennium and Millennium-II simulations \citep{Mill05_short,MillII09}
to the distribution of galaxy stellar masses found by \cite{LiWh09}
using Sloan Digital Sky Survey data~\citep{SDSS7_short}.
\cite{Guea10} found that the ratio of total stellar mass $M_\ast$ to
halo mass $M_{v}$ was well fit by
\begin{equation} \label{eq:mast} M_\ast = M_{v} \times
  A\left[\left(\frac{M_{v}}{M_0}\right)^{-\alpha} +
    \left(\frac{M_{v}}{M_0}\right)^{\beta}\right]^{-\gamma},
\end{equation}
with an intrinsic scatter of order $0.2$ in $\log_{10} M_\ast$, where
$A=0.129$, $M_0 = 10^{11.4}\msun$, $\alpha=0.926$, $\beta=0.261$ and
$\gamma=2.440$.

To gain insight into the likely value of the concentration parameter
$c_{v^\prime}$ we consider the study of \cite{BKea10}. This examined
haloes taken from the Millennium-II simulations at redshift zero, in
the mass range $10^{11.5}\leq M_v/[h^{-1}\msun]\leq10^{12.5}$ -- a
mass range that the Milky Way's halo is likely to lie in -- and
determined that the probability distribution of the concentration was
well fit by a Gaussian distribution in $\ln c_{v^\prime}$, with
\begin{equation}
  \ln c_{v^\prime} = 2.256 \pm 0.272.
\end{equation}
Cosmological simulations predict that typical halo concentration
varies with halo mass, but typically only weakly
\citep[e.g. $c_{v}\propto M_{v}^{-0.1}$,][]{Neea07} when compared to
the intrinsic scatter in $c$ at a given virial mass, so we accept the
relationship given by \cite{BKea10} as stated, independent of
$M_{v^\prime}$. Baryonic physics is likely to have an effect on the
concentration, much as it does on the inner density profile
\citep{Duea10}, but we neglect that in this study.

The Galaxy's dark-matter halo is far more massive than its stellar
halo, which has a mass $\sim4\times10^8\msun$ \citep{Beea08_short}. We
therefore treat the stellar halo as a negligible fraction of the
Galactic halo, and do not consider it further.

\section{Kinematic data} \label{sec:data}

\subsection{The Solar position and velocity}
\label{sec:sun} 
If we do not know the position and velocity of the Sun it is extremely
difficult to interpret any other kinematic observations of the Milky
Way.  Unfortunately there is still significant uncertainty on the key
parameters, namely the distance from the Sun to the Galactic Centre
$R_0$, the rotational speed of the local standard of rest (LSR) 
$v_0$, and the velocity of the Sun with respect to the LSR $\vsol$ \citep[see
e.g.][]{PJMJJB10:masers}.  In this work we take $R_0$ from the study
of stellar orbits around the supermassive black hole at the Galactic
Centre, Sgr A*, by \cite{Giea09}
\begin{equation} \label{eq:R0} R_0 = 8.33\pm0.35\kpc ,
\end{equation}
though we could equally have taken the value $8.4\pm 0.4\kpc$
from \cite{Ghea08}, found from similar data.

A series of recent papers \citep*{JJB10,PJMJJB10:masers,SBD10}
have re-examined evidence regarding the value of $\vsol$.
We use the value determined by \citeauthor*{SBD10}:
\begin{equation}\label{eq:vsol}
  \begin{array}{lll}
    \vsol & = &  (U_\odot,V_\odot,W_\odot) \\
    & = & (11.1,12.24,7.25)\pm(1,2,0.5) \kms 
  \end{array}
  ,
\end{equation}
where $U_\odot$ is the velocity towards the Galactic Centre, $V_\odot$
is the velocity in the direction of Galactic rotation, $W_\odot$ is
the velocity perpendicular to the Galactic plane. We quote the
systematic uncertainties here, because these dominate the total
uncertainty (especially in $V_\odot$).

Many studies that can be used to constrain the value of $v_0$ are
ones which actually constrain the ratio $v_0/R_0$ -- for example using
the Oort constants $A$ and $B$, which can be determined locally
\citep[e.g.][]{FeWh97} where $A-B = v_0/R_0$. In this study, we use
the proper motion of Sgr A* in the plane of the Galaxy, as determined
by \cite{ReBr04}
\begin{equation} \label{eq:sgra} \mu_{\mathrm{Sgr A^\ast}} = -6.379\pm 0.026\,
  \mathrm{mas}\,\mathrm{yr}^{-1}.
\end{equation}
Since Sgr A* is expected to be fixed at the Galactic Centre to within
$\sim1\kms$, this proper motion is thought to be almost entirely due
to the motion of the Sun around the Galactic Centre
$(v_0+V_\odot)/R_0$.  This measurement is sufficiently accurate (and
the assumed velocity of Sgr A* sufficiently small) that the greatest
uncertainty in the value of $v_0/R_0$ required for our models actually
comes from the uncertainly in the value of $V_\odot$!

This measured proper motion is inconsistent with the value of $A-B$ 
found by~\cite{FeWh97} -- who give the most commonly used values
for $A$ and $B$.  Therefore we do not attempt to use the Oort
constants as constraints.

\subsection{Terminal velocity curves} \label{sec:vt} For circular
orbits in an axisymmetric potential, the peak velocity of the
interstellar medium (ISM) along any line-of-sight at Galactic
coordinates $b=0$ for $-90<l<90$ corresponds to the gas at the
tangent point, at Galactocentric radius $R = R_0\sin{l}$. This is known
as the terminal velocity, and is given by
\begin{equation}
  \label{eq:vterm}
  v_{\mathrm{term}} = v_\mathrm{c}(R_0\;\sin{l}) - v_\mathrm{c}(R_0)\sin{l}.
\end{equation}
\cite{Ma94,Ma95} produced data which gave this peak velocity for a
range of lines of sight in the Galaxy from a number of surveys
\citep{WeWi74,Keea86,BaLo84,Knea85}.

To constrain our model, we have to take account of the
effects that non-axisymmetric structure in the Galaxy and non-circular
motion of the ISM will have on these data. 
To do so we follow DB98 in allowing for an uncertainty of $7\kms$ in
$v_{\mathrm{term}}$ at any given $l$, and 
restricting ourselves to
data at $|\sin l|>0.5$, which is unlikely to be significantly affected
by the bar.

\subsection{Maser observations} \label{sec:maser} In recent years it
has become possible to use Galactic maser sources as targets for
astrometric measurements that are sufficiently accurate for
parallaxes with uncertainties $\sim10\mu\mathrm{as}$ to be determined for them
\citep[e.g.][]{Reea09_short}.  \cite{PJMJJB10:masers} showed that these
observations were consistent with models that placed the masers on
near circular orbits with $v_0/R_0$ similar to the value implied by
the proper motion of Sgr A* (Section~\ref{sec:sun}).  However
$R_0$ by itself (or, equivalently, $v_0$) was shown to depend quite
strongly on the shape of the rotation curve. We constrain our mass
model using these data by applying a version of the likelihood
analysis used by \cite{PJMJJB10:masers}.

The likelihood analysis for these data requires integrating over
heliocentric radius the probability density of the observations
(parallax, proper motion and radial velocity) given a model for the
expected velocity distribution.  The maser sources are associated with
high mass star formation regions, and are expected to be on
near-circular orbits. We model the velocity distribution as circular
rotation (with a velocity which depends on the potential), plus a
random component that has a Gaussian probability distribution, uniform
in direction, with width $\Delta_v$. The interested reader can find
details in \cite{PJMJJB10:masers}. However the analysis in this study differs
from that of \cite{PJMJJB10:masers} in that (a) the rotation curve, including $v_0$, is defined by the
mass model; (b) the value of $\vsol$ is assumed to be that in
eq. (\ref{eq:vsol}); (c) we include data from~\cite{Ryea10} and
\cite{Saea10}, which were not available when the~\cite{PJMJJB10:masers}
study was performed; (d) in the interests of simplicity we fix
$\Delta_v=7\kms$, close to the best-fitting values found when
$\Delta_v$ was allowed to vary in the previous study, and (e) we
assume the masers have no \emph{systematic} velocity offset from the
rotation curve -- one of the conclusions of \cite{PJMJJB10:masers} was
that no such offset is required to explain the data.

\subsection{Vertical force} \label{sec:kz}
\cite{KuGi91} used observations of K stars to find the vertical force
at $1.1\kpc$ above the plane at the Solar radius, $K_{z,1.1}$. We
adopt the value they found
\begin{equation}
  K_{z,1.1} = 2\pi G \times (71\pm6)\msun\pc^{-2}
\end{equation}
as a constraint.

\subsection{Observational constraints on the structure at large radii}
\label{sec:larger}

\begin{table*}
  \begin{tabular}{c|c|c|c}
    Property constrained & Constraint & Section described & Source \\ \hline
Bulge profile & see equation (\ref{eq:bulge}) &\ref{sec:bulge} & \cite{BiGe02} \\
$M_\mathrm{b}$ & $(8.9\pm0.89)\times 10^9\msun$ & \ref{sec:bulge} & \cite{BiGe02} \\
Disc profile & Double exponential & \ref{sec:disc} & - \\ 
$z_{\mathrm{d},\mathrm{thin}}$ & $0.3\kpc$ & \ref{sec:disc} & \cite{Juea08_short} \\
$z_{\mathrm{d},\mathrm{thick}}$ & $0.9\kpc$  & \ref{sec:disc} & \cite{Juea08_short} \\
$R_{\mathrm{d},\mathrm{thin}}$&$2.6\pm0.52\kpc$& \ref{sec:disc} & \cite{Juea08_short} \\
$R_{\mathrm{d},\mathrm{thick}}$&$3.6\pm0.72\kpc$ & \ref{sec:disc} & \cite{Juea08_short} \\
$f_{\mathrm{d},\odot}$ & $0.12\pm0.012$ & \ref{sec:disc} & \cite{Juea08_short} \\
Halo profile & NFW profile & \ref{sec:halo} & \cite{NFW96} \\
$M_\ast/M_v$ & see equation (\ref{eq:mast}) & \ref{sec:halo} & \cite{LiWh09} \\
$\ln c_{v^\prime}$ & $2.256\pm0.272$ &  \ref{sec:halo} & \cite{BKea10} \\
$R_0$ & $8.33\pm0.35\kpc$ & \ref{sec:sun} & \cite{Giea09} \\
$\mu_{\mathrm{Sgr A^\ast}}$ & $-6.379\pm 0.026\,\mathrm{mas}\,\mathrm{yr}^{-1}$ &
 \ref{sec:sun} & \cite{ReBr04} \\
$K_{z,1.1} $ & $2\pi G \times (71\pm6)\msun\pc^{-2}$ & \ref{sec:kz} & \cite{KuGi91} \\
$M_{50}$ & $\lesssim 5.4\times10^{11}\msun$, see eq. (\ref{eq:m50}) & \ref{sec:larger} & \cite{WiEv99} \\ \hline
\multicolumn{2}{c}{Kinematic data} &  Section described & Source \\ \hline
\multicolumn{2}{c}{Terminal velocities} & \ref{sec:vt} & \cite{Ma94,Ma95} \\
\multicolumn{2}{c}{\multirow{2}{*}{Maser observations}} & 
\multirow{2}{*}{\ref{sec:maser}} & \cite{Reea09_short,Ryea10}; \\ & & & \cite{Saea10} \\
  \end{tabular}
  \caption{
    Summary of all the constraints applied to the models.
  }
  \label{tab:constr}
\end{table*}  

It is extremely difficult to gain useful constraints on the structure
of the Milky Way at large radii from observational data. Any
population of dynamical tracers suffers from small
number statistics and/or poor observational constraints
(especially on proper motion) and because of uncertainty over the
associated distribution function, especially the velocity anisotropy,
and whether or not all of the population is in fact bound to the Milky
Way. This leads to total halo mass estimates which have fractional
uncertainties of order unity, e.g. $1.9^{+3.6}_{-1.7}\times
10^{12}\msun$ \citep{WiEv99},
$0.3$ to $2.5\times10^{12}\msun$ \citep{Baea06}.

It is possible to take cosmologically motivated simulations and use
them to provide some insight into the expected distribution function
of a tracer population. \cite{Xuea08_short} did this for a sample of blue
horizontal branch (BHB) stars and found that the mass at radii less
than $60\kpc$, under this assumption, was $M(<60\kpc) =
(4.0\pm0.7)\times10^{11}\msun$. However this value is entirely
dependent on the assumption that these cosmological simulations (and
their prescriptions for star-formation, feedback etc.) can be relied
upon to predict accurately the distribution function of BHB stars.

Another possible source of information is the motion of the Magellanic
clouds, but this can only be used straightforwardly if they can be
assumed to be bound to the Milky Way -- \cite{Beea07} have shown that
this may not be a valid assumption.

Other authors \citep[notably][]{Smea07_short} have attempted to determine
the local escape velocity from the Milky Way, and use this to
constrain the mass distribution at large radii. However this is
determined under the assumption that the population of stars close to
the escape velocity can be described by a steady-state distribution
function that is depends on energy alone. Stars that have velocities
that lie close to the escape velocity are on orbits with very long
periods, which means that there is no reason to expect them to be in a
steady state, especially since it is likely that a large fraction of
halo stars are associated with
streams. 
The assumed distribution function is isotropic (because it only
depends on energy), and this assumption is unlikely to be
valid --
halo stars in cosmological simulations typically have a distribution
function which is radially anisotropic. 

The ``Timing Argument'' has been used to constrain the mass of the
local group or that of the Milky Way itself. In its simplest form, this
assumes that the currently observed radial velocity of M31 towards the
Milky Way must be due to the gravitational interaction of the two
galaxies overcoming the overall cosmic expansion, and that the
corresponding orbital motion must have happened within the age of the
universe, and shows that this implies a lower limit on the mass of the
two galaxies. \cite{LiWh08} extended this argument by searching cosmological 
simulations for
comparable galaxy pairs that were moving towards one another in the
simulations. They also looked for galaxy pairs comparable to the Milky
Way and Leo I, which are moving away from one another, and used the
Timing Argument under the assumption that Leo I is moving towards
apocentre for a second time. This yielded a probability distribution
on $M_{v}$ with lower and upper quartiles
$[1.78,3.09]\times10^{12}\msun$ and a $95\%$ confidence lower limit of
$7.94\times10^{11}\msun$.  This approach rather relies upon Leo I
being bound to the Milky Way, which it may not be, and the authors
note that, compared to the Milky Way/M31 system, the ``complex
dynamical situation offers greater scope for uncertainty''.

While \cite{WiEv99} quoted a very uncertain figure for the total halo
mass, they also found that the mass within $50\kpc$ was a more
robustly determined quantity: $M_{50} =
5.4^{+0.2}_{-3.2}\times10^{11}\msun$. The quoted uncertainty is very
asymmetrically distributed about the most likely value, and we choose
to see this result as providing an upper bound on the mass inside
$50\kpc$, which we take into account in our analysis through the
probability distribution
\begin{equation}\label{eq:m50}
  P(M_{50})= \left\{ 
    \begin{array}{ll}
      C & \mathrm{for }\,M_{50} \leq M_{\mathrm{WE}} \\
      C\exp\left(-\left[\frac{M_{50}-M_{\mathrm{WE}}}{\delta_{M_{50}}}
        \right]^2\right) &\mathrm{for }\,  M_{50} > M_{\mathrm{WE}} \\
    \end{array}
  \right.
\end{equation}
where $M_{\mathrm{WE}}=5.4\times10^{11}\msun$,
$\delta_{M_{50}}=2\times10^{10}\msun$, and $C$ is a normalisation
constant. The only effect of using this $P(M_{50})$ constraint is to
penalise models with $M_{50}>M_{\mathrm{WE}}$. For models with
$M_{50}\ll M_{\mathrm{WE}}$ this would be an inappropriate choice but,
as we show in Section~\ref{sec:res}, that is not the case for this
study.

Given the difficulties described here, we have decided not to
constrain our model with the results of any other study. We do compare
our best-fitting model to them (Section~\ref{sec:compare})

\section{Fitting the models} \label{sec:fit}
We use Bayesian statistics to find the probability density function
(pdf) for our model parameters given the kinematic data described in
Section~\ref{sec:data}, and the prior probabilities given in
Section~\ref{sec:comp} (and the value of $R_0$, given in
Section~\ref{sec:data}). We refer to the parameters collectively as
$\bolth$, and the data as $d$. Bayes theorem tells us that this pdf is
then
\begin{equation}
  p(\bolth|d) = \frac{\mathcal{L}(d|\bolth)\,p(\bolth)}{p(d)}
\end{equation}
where the total likelihood $\mathcal{L}(d|\bolth)$ is the product of
the likelihoods associated with each kinematic data-set or constraint
described in Section~\ref{sec:data}, and $p(\bolth)$ is the
probability of a parameter set given the prior probability
distributions described in Section~\ref{sec:comp} (and that on
$R_0$). The Bayesian Evidence, $p(d)$, is often a very important
quantity, but in this study it is an unimportant normalisation
constant, so we ignore it.

Given the constraints we have applied to the components described in
Section~\ref{sec:comp}, we are left with 8 model parameters that we
allow to vary: The scale-lengths and density normalisations of the
thin and thick discs
($R_{\mathrm{d},\mathrm{thin}},\Sigma_{\mathrm{d},0,\mathrm{thin}},R_{\mathrm{d},\mathrm{thick}},\Sigma_{\mathrm{d},0,\mathrm{thick}}$); the
density normalisation -- and thus mass -- of the bulge ($\rho_{\mathrm{b},0}$);
the scale-length and density normalisation of the CDM halo
($r_\mathrm{h},\rho_{\mathrm{h},0}$); and the solar radius ($R_0$). While each of these
parameters is free to vary, each one is explicitly or implicitly
associated with at least one prior probability distribution.

To explore the pdf $p(\bolth|d)$ we use the Metropolis algorithm
\citep{Metropolis}, which is a Markov Chain Monte Carlo method for
drawing a representative sample from a probability distribution. This
allows us both to find the peak of the pdf (to reasonable accuracy)
and to characterise its shape.  We start with some choice for the
parameters $\bolth_n$, and calculate $p(\bolth_n|d)$. We then
\begin{enumerate}
\item choose a trial parameter set $\bolth '$ by moving from
  $\bolth_n$ in all directions in parameter space, by an amount chosen
  at random from a ``proposal density'' $Q(\bolth ',\bolth_n)$ (see
  below);
\item determine $p(\bolth '|d)$;
\item choose a random variable $r$ from a uniform distribution in the
  range [0,1];
\item if $p(\bolth '|d)/p(\bolth_n|d) > r$, accept the trial parameter
  set, and set $\bolth_{n+1} = \bolth '$. Otherwise set
  $\bolth_{n+1}=\bolth_n$.
\item Return to step (i), replacing $\bolth_n$ with $\bolth_{n+1}$.
\end{enumerate}
The first few values of $\bolth$ are ignored as ``burn-in'', which
helps to remove the dependence on the initial value of $\bolth$.

The pdf is quite narrow in some directions in the parameter space, but
these directions are not simply parallel to the coordinate
axes. Therefore it is efficient to use a proposal density which is
aligned (to a reasonable approximation) with the pdf -- this is
allowed by the Metropolis algorithm as the only constraints on
$Q(\bolth_1,\bolth_2)$ are that it is symmetrical with respect to
swapping $\bolth_1$ and $\bolth_2$, and makes it possible to reach any 
point in phase space. We therefore use the Metropolis
algorithm in two phases -- first with a proposal density which is
simply made up of Gaussian distributions in each parameter
individually, with associated step size chosen by hand. The resultant
chain (of a relatively short length, and after a burn-in) is then used
to construct a covariance matrix that is then used to define a new
$Q(\bolth ',\bolth_n)$. This new proposal density is, again, a
multi-variate Gaussian but with the principal axes now aligned to the
eigenvectors of the covariance matrix, and with the step size in each
direction related to the respective eigenvalues. The chain produced
with this second proposal density is then used in all calculations.

\section{Results}\label{sec:res}

\begin{figure}
  \centerline{\resizebox{\hsize}{!}{\includegraphics{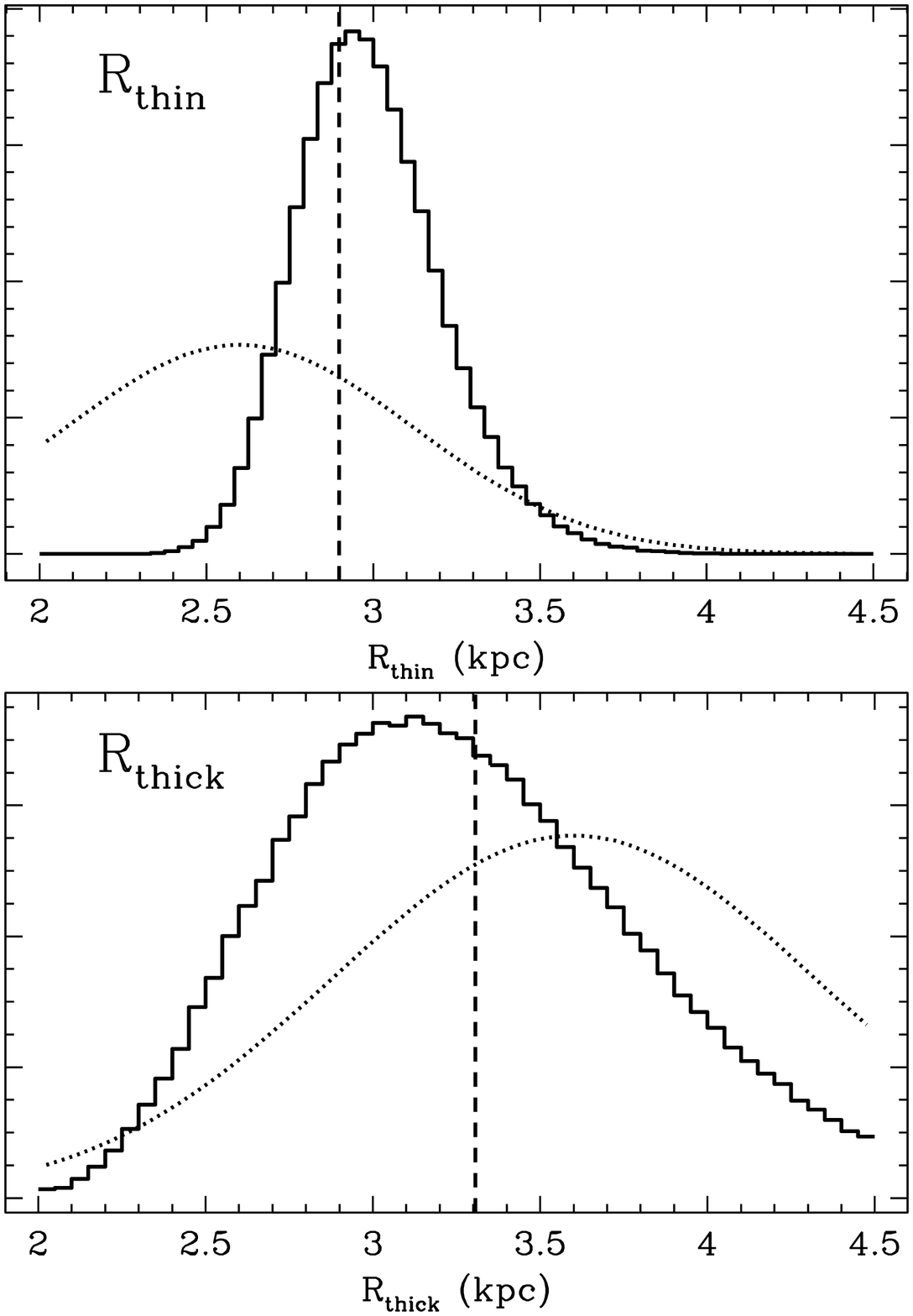}}}
  \caption{Histogram of the pdf of the thin (top) and thick (bottom) disc 
    scale-lengths in our models (solid histogram) normalised over all other
    parameters, compared in each case to the prior 
    pdf described in Section~\ref{sec:disc} (dotted).
    In each plot the value corresponding to our best-fitting model is shown
    as a dashed vertical line. In each case the posterior pdf is peaked 
    near to $3\kpc$.
\label{fig:Rdisc}
}
\end{figure}

\begin{figure}
  \centerline{\resizebox{\hsize}{!}{\includegraphics{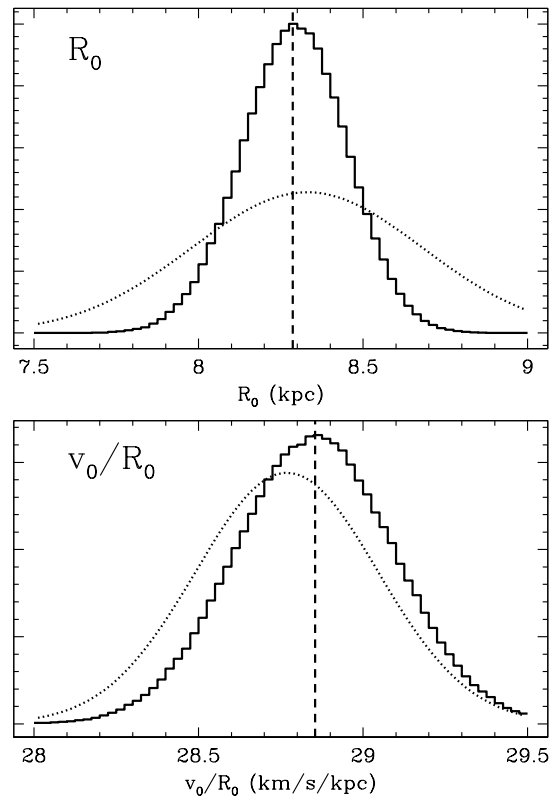}}}
  \caption{Histogram of the pdf of $R_0$ (top) and $v_0/R_0$ (bottom)
    normalised over all parameters (solid line). The prior pdf
    on $R_0$ and the pdf on $v_0/R_0$ associated with the apparent 
    motion of Sgr A* (Section~\ref{sec:sun}) are plotted as dotted 
    lines. As in Figure~\ref{fig:Rdisc}
    the value corresponding to our best-fitting model is shown
    as a dashed vertical line. 
\label{fig:Solar}
}
\end{figure}

\begin{figure}
  \centerline{\resizebox{\hsize}{!}{\includegraphics{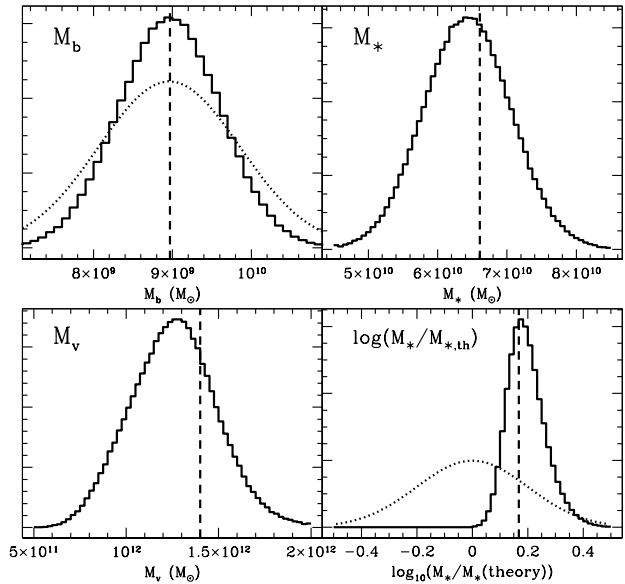}}}
  \caption{Histogram of the pdf for the bulge mass $M_\mathrm{b}$ (top-left), the total
    stellar mass $M_\ast$ (top-right), the virial mass $M_{v}$ (bottom-left) 
    and the log of the ratio of $M_\ast$ to the value of $M_\ast$ predicted by
    equation (\ref{eq:mast}) (bottom-right). In first and last case, 
    where a straightforward 
    prior applies to the quantity shown, it is plotted as a dotted curve.
    Again, the value corresponding to our best-fitting model is shown
    as a dashed vertical line in each plot. 
\label{fig:Mass}
}
\end{figure}

\begin{table*}
  \begin{tabular}{ccccccccccc} 
    & & $\Sigma_{\mathrm{d},0,\mathrm{thin}}$ & $R_{\mathrm{d},\mathrm{thin}}$ & $\Sigma_{\mathrm{d},0,\mathrm{thick}}$ & $R_{\mathrm{d},\mathrm{thick}} $ & 
    $\rho_{\mathrm{b},0}$ & $\rho_{\mathrm{h},0}$ & $r_\mathrm{h} $ & $R_0 $ & \\ \hline
    Best & & 816.6 & 2.90 & 209.5 & 3.31 & 95.6 & 0.00846 & 20.2 & 8.29 &\\ 
    Convenient & & 753.0 & 3 & 182.0 & 3.5 & 94.1 & 0.0125 & 17 & 8.5 &\\ \hline
    Mean & & $741$ & $3.00$ & $238$ & $3.29$ & $95.5$ & $0.012$ & $18.0$ & 
    $8.29$ & \\
    Std. Dev. & &123 & 0.22 & 110 & 0.56 & 6.9 & 0.006 & 4.3 & 0.16 &  \\ \hline
    & $v_0$ & $M_{\mathrm{b}}$ & $M_\ast$  & $M_v $ & $K_{z,1.1}$& $\Sigma_{\mathrm{d},\odot}$ & 
    $\rho_{\ast,\odot}$ & $\rho_{\mathrm{h},\odot}$ & $f_{\mathrm{d},\odot}$ &
    $g_{\ast,\odot}/g_{\mathrm{h},\odot}$ \\ \hline
    Best & 239.1 & $8.97$ & $66.1$ & $1400$ & 77.7 & 63.9 & 0.087 & 0.0104 & 
    0.122 & 1.38 \\
    Convenient & 244.5 &    8.84 & 65.1 & 1340 & 75.4 & 60.3 & 0.083 & 
    0.0111 & 0.121 & 1.12 \\\hline
    Mean &  $239.2$ & $8.96$ & $64.3$ & $1260$ & 76.5 & 62.0 & 0.085 & 0.0106 & 0.120 & 1.29 \\
    Std. Dev.  & 4.8 & 0.65 & 6.3 & 240 & 5.3 & 7.6 & 0.010 & 0.0010 & 0.012 & 0.30 \\
  \end{tabular}
  \caption{
    Parameters (upper) and derived properties (lower) 
    of our best-fitting model, our ``convenient'' model, 
    and mean and standard deviation marginalised over all the (other) parameters
    in the pdf. The density profile of the bulge follows 
    the description in eq. (\ref{eq:bulge}), with other parameters as
    described directly below that equation; the density profiles 
    of the discs follow     eq. (\ref{eq:disc}), with scale-heights 
    $0.3$ and $0.9\kpc$ for thin and thick discs respectively;  and the halo
    density profile is that given in eq. (\ref{eq:NFW}). $\Sigma_{\mathrm{d},\odot}$ is 
    the disc surface density at the Sun; $\rho_{\ast,\odot}$ and 
    $\rho_{\mathrm{h},\odot}$ are the local 
    densities of the stellar component and CDM 
    halo, respectively; $g_{\ast,\odot}/g_{\mathrm{h},\odot}$ is the ratio of the 
    gravitational force on the Sun from the stellar component ($g_{\ast,\odot}$)
    to that from the CDM halo ($g_{\mathrm{h},\odot}$) -- this is a measure of 
    whether the Galaxy is disc- or halo-dominated. The means and
    standard deviations of the parameters are not always particularly helpful
    statistics, as they say nothing about the correlations between parameters.
    It should also be noted that the true uncertainties on 
    $\Sigma_{\mathrm{d},0,\mathrm{thin}}$, $\Sigma_{\mathrm{d},0,\mathrm{thick}}$
    and $\rho_{\ast,\odot}$ are considerably larger than the standard 
    deviations quoted here, because the disc scale heights have been held 
    constant (see Section~\ref{sec:height}). 
    Distances are quoted in units of $\kpc$, velocity in $\kms$, 
    masses in $10^9\msun$, surface densities in $\msun\pc^{-2}$,
    densities in $\msun\pc^{-3}$, and $K_{z,1.1}$ in units of 
    $(2\pi G)\times\msun\pc^{-2}$.  
  }
  \label{tab:param}
\end{table*}

\begin{table*}
  \begin{tabular}{c|cccccccc} 
    & $\Sigma_{\mathrm{d},0,\mathrm{thin}}$ & $R_{\mathrm{d},\mathrm{thin}}$ & $\Sigma_{\mathrm{d},0,\mathrm{thick}}$ & $R_{\mathrm{d},\mathrm{thick}} $ & 
    $\rho_{\mathrm{b},0}$ & $\rho_{\mathrm{h},0}$ & $r_\mathrm{h} $ & $R_0 $ \\ \hline
    $\Sigma_{\mathrm{d},0,\mathrm{thin}}$ & 1 &  &  &  &  &  &  &  \\
    $R_{\mathrm{d},\mathrm{thin}}$  & $-0.727$ & 1 &  &  &  &  &  &  \\
    $\Sigma_{\mathrm{d},0,\mathrm{thick}}$ & $-0.467$ & 0.483 & 1 &  &  &  &  &  \\
    $R_{\mathrm{d},\mathrm{thick}} $  & 0.468 & $-0.304$ & $-0.845$ & 1 &  &  &  &  \\
    $\rho_{\mathrm{b},0}$  & $-0.197$ & 0.167 & $-0.047$ & 0.037 & 1 &  &  &  \\
    $\rho_{\mathrm{h},0}$  & $-0.632$ & 0.328 & $-0.247$ & 0.151 & $-0.009$ & 1 &  &  \\
    $r_\mathrm{h} $  & 0.588 & $-0.299$ & 0.221 & $-0.130$ & 0.003 & $-0.877$ & 1 &  \\
    $R_0 $  & $-0.369$ & 0.602 & $-0.071$ & 0.157 & $-0.017$ & 0.373 & $-0.338$ & 1 \\

  \end{tabular}
  \caption{Correlation matrix for the model parameters. A value of 1 corresponds 
    to perfect correlation (e.g. any parameter with itself), $-1$ corresponds to
    perfect anti-correlation, 0 to no correlation. The strongest relationships
    are anti-correlations
    between parameters which, taken in combination, define the mass of a given 
    component or its density at a given point -- i.e. between 
    $\rho_{\mathrm{h},0}$ \& $r_\mathrm{h} $, between $\Sigma_{\mathrm{d},0,\mathrm{thin}}$ \& 
    $R_{\mathrm{d},\mathrm{thin}}$ and between $\Sigma_{\mathrm{d},0,\mathrm{thick}}$ \& $R_{\mathrm{d},\mathrm{thick}}$.}
  \label{tab:corr}
\end{table*}

In Figures~\ref{fig:Rdisc} to \ref{fig:Mass} we plot the probability
density functions of various quantities associated with our models,
marginalised over all parameters, as determined by the Metropolis
algorithm. Where appropriate we also plot the prior probability
distribution directly associated with each.  The value associated with
our best-fitting model is indicated on each plot with a dashed
vertical line. In Table~\ref{tab:param} we give the parameters of 
our best-fitting model, and those of what we call the ``convenient''
model (see below), along with the mean and standard deviation for each
parameter, marginalised over other parameters in the pdf.

The marginalised distributions do not give a sense of the correlations
between parameters, so in Table~\ref{tab:corr} we show the correlation
matrix of the various parameters $\bolth$. For the $i$,$j$th
component, $\mathrm{corr}(\theta_i,\theta_j)$, this takes the value
\begin{equation}
  \mathrm{corr}(\theta_i,\theta_j) = \frac{\mathrm{cov}(\theta_i,\theta_j)}
  {\sigma_i\sigma_j}
\end{equation}
where $\mathrm{cov}(\theta_i,\theta_j)$ is the covariance. A value of
1 corresponds to a perfect correlation, $-1$ corresponds to a perfect
anti-correlation, and 0 corresponds to no correlation. The
correlation matrix is manifestly symmetric, so in Table~\ref{tab:corr} 
we only show half of it.

The strongest correlations or anti-correlations are between parameter
pairs that describe a single component -- this explains, for example,
why the spread in $M_\ast$ is so much smaller than the spread in
$\Sigma_{\mathrm{d},0,\mathrm{thin}}$ or $\Sigma_{\mathrm{d},0,\mathrm{thick}}$, and why the standard
deviation of $\rho_{\mathrm{h},0}$ can be as large as $\sim50\%$ while that in
$M_v$ is $\sim20\%$ (and that in $M_{50}$ is even smaller, see
below). Other fairly strong correlations are noticeable between
$\Sigma_{\mathrm{d},0,\mathrm{thin}}$ and virtually every other parameter, and between
$R_{\mathrm{d},\mathrm{thin}}$ and $R_0$.

Figure~\ref{fig:Rdisc} shows the pdfs associated with the disc scale-lengths, 
$R_{\mathrm{thin}}$ and $R_{\mathrm{thick}}$. Both have peaks around $3\kpc$,
and the best-fitting values also lie quite near $3\kpc$. That the
kinematic data (primarily the terminal velocity curve) drives the two
scale-lengths towards a similar value is unsurprising -- the two discs
have effects on the dynamics of the Galaxy that scale very similarly. 
The posterior pdf on $R_{\mathrm{thin}}$ is much narrower than the prior, which
is not true to the same extent for $R_{\mathrm{thick}}$; this is not surprising
given that the thin disc is more massive than the thick disc so the
constraints based on kinematic data are more likely to dominate the photometric
constraints in the thin disc pdf.

DB98 considered mass models with four different scale-lengths between
$2\kpc$ and $3.2\kpc$, and halo density profiles which could take on a
wide range of power-law slopes in their inner and outer parts. They
found that the models with disc scale-lengths of $2.4\kpc$ or smaller
required haloes which, unrealistically, had density profiles which
\emph{increased} with radius ($\rho\propto r^2$) in their inner parts,
whereas the models with scale-lengths $2.8\kpc$ or greater did
not. Therefore, given that we require a halo profile that increases
as $\rho\propto r^{-1}$ in its centre, we should not be surprised that
this favours models with disc scale-lengths larger than $2.5\kpc$.

Figure~\ref{fig:Solar} shows the posterior pdfs for the position of the Sun
and the circular speed at the Sun in our models. The posterior pdf on $R_0$ is
much narrower than the prior, with a standard deviation of $0.16\kpc$,
compared to an uncertainty on the prior of $0.33\kpc$. The posterior
pdf on $v_0/R_0$ is of a similar width to that which would be seen if
our only information on its value was the proper motion of Sgr A*, but
it is slightly displaced to higher values of $v_0/R_0$. This
displacement is most likely due to the maser data
(Section~\ref{sec:maser}), which~\cite{PJMJJB10:masers} showed are best
fit by models in which $v_0/R_0$ is slightly larger than the value
implied by equation (\ref{eq:sgra}). The corresponding value of $v_0$, 
$239.2\pm4.8\kms$, has a significantly smaller uncertainty than 
simply combining the \cite{Giea09} value of $R_0$ and the 
\cite{ReBr04} value of $\mu_{\mathrm{Sgr A^\ast}}$ gives, primarily 
because of the tighter constraints on $R_0$.

The posterior pdf on bulge masses (Figure~\ref{fig:Mass}, top left) is
close to our prior from \cite{BiGe02}.
The bottom right panel of 
Figure~\ref{fig:Mass} shows that the stellar mass of the
models is tightly peaked at values towards the high end of those we
would expect given the virial mass, and our constraint on this
relationship given in equation (\ref{eq:mast}).  This can equivalently be 
thought of as corresponding to low halo masses, given the stellar
masses. Compared to the prior pdfs taken from theoretical
considerations (Section~\ref{sec:halo}), the haloes have low masses
and high concentrations ($\ln c_{v^\prime}=2.73$ for the best-fitting model, 
and with mean and standard deviation $2.83$ and $0.18$
respectively). The halo is likely to be sub-dominant at $R_0$ in the sense 
that the gravitational force on the Sun is mostly from the stellar 
component, though this is not the case for all models (ratio of gravitational
forces $g_{\ast,\odot}/g_{\mathrm{h},\odot} = 1.29\pm0.30$).
 The mass inside $50\kpc$ of our best-fitting model is
$5.3\times10^{11}\msun$, with mean and standard deviation
$5.1\pm0.4\times 10^{11}\msun$. This is close to the mass limit we
take from \cite{WiEv99}, indicating that the latter is an important upper
limit on our models.

The local density normalisation $f_{\mathrm{d},\odot}$ 
is $0.122$ for our best-fitting
model. The mean and standard deviation are $0.120\pm0.012$, which is
exactly what one would expect from our prior alone. 
This is due to the fact that our kinematic
data does not provide us with any great ability to discriminate between
mass in either of the two discs. 
Therefore, the value of 
$f_{\mathrm{d},\odot}$ almost entirely defines (with the disc scale-heights)
our constraints on the relative contributions of the thick and thin disc.

\cite{CaUl10} used a rather similar method to the one described in
this paper with the single intention of determining the local dark
matter density, finding a value $\sim0.39\Gevdens$. We have made a
number of different assumptions, and taken into account different
constraints, but many of the key assumptions (axisymmetry, exponential
disc and a halo profile motivated by CDM simulations) are the same,
and we find a similar local dark matter density $0.40\pm0.04\Gevdens$. It is
well worth noting that the uncertainty we quote here is purely the
statistical uncertainty associated with this set of parametrized
models, and the constraints we have chosen to apply. In particular, we
have made the approximations that the discs are well described by
exponential profiles at all radii, and that the halo density profile
is well described by a spherically symmetric NFW profile. These
assumptions and others, including ones which are not not directly
related to the dark matter profile, will have a significant effect on
the value found for the local dark matter density.  We expect that the
systematic uncertainties on this value are much larger than the
statistical uncertainty we state here.

It is clear from Figures~\ref{fig:Rdisc} to \ref{fig:Mass} that there
is a wide range of models that fit our constraints almost as well as
our best-fitting model. It is often convenient to use models in which
certain key parameters are chosen to take simple values (for example it
is common to take $R_0=8$ or $8.5\kpc$), and we provide such a model
in Table~\ref{tab:param}.  To take account of the correlations between
parameters we chose to build this ``convenient'' model one step at a
time. We first hold $R_0$ constant at a suitable value determined from
Figure~\ref{fig:Solar}, and find the pdf associated with these models,
which we use to choose a suitable thin disc scale length
$R_{\mathrm{d},\mathrm{thin}}$. We then repeat this process to find a suitable value for
$R_{\mathrm{d},\mathrm{thick}}$, and then again for $r_\mathrm{h}$. Following this procedure we
take $R_0=8.5\kpc$, $R_{\mathrm{d},\mathrm{thin}}=3\kpc$, $R_{\mathrm{d},\mathrm{thick}}=3.5\kpc$ and
$r_\mathrm{h}=17\kpc$.

\begin{figure}
  \centerline{\resizebox{\hsize}{!}{\includegraphics[angle=270]{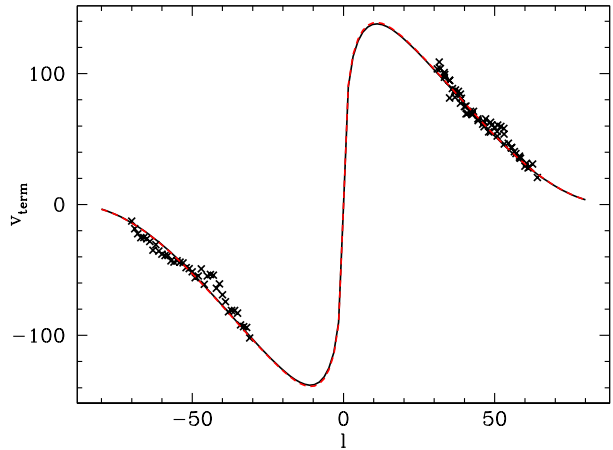}}}
  \caption{ Comparison of the terminal velocity curve predicted by our
    best-fitting model (solid curve), the ``convenient'' model (dashed red 
    curve) and the terminal velocity measurements from the Galaxy (crosses). 
    The two model curves are nearly indistinguishable.
    \label{fig:vt}
  }
\end{figure}

\begin{figure}
  \centerline{\resizebox{\hsize}{!}{\includegraphics[angle=270]{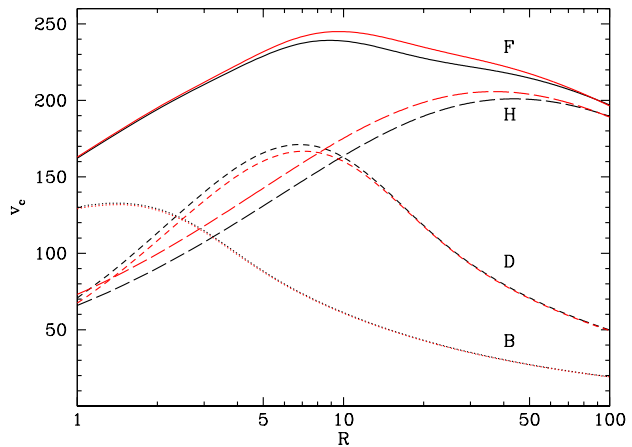}}}
  \caption{ The rotation curve for our best-fitting (black) and convenient 
    (red) models -- note that radius is plotted logarithmically. 
    The solid line, labelled F, is the full rotation curve, 
    with the other 
    curves showing, in each case, the contribution of the bulge (B, dotted), 
    discs
    (D, short-dashed) and halo (H, long-dashed). 
    \label{fig:rc}
  }
\end{figure}

In Figure~\ref{fig:vt} we compare the terminal velocity curve
associated with our best-fitting model, and with the ``convenient''
model, to the observational data (Section~\ref{sec:vt}). The
difference between the curves of the two models is very
small, and both provide a good fit to these data. In
Figure~\ref{fig:rc} we plot the rotation curves of the best-fitting
and convenient models, and the decomposition into the components due
to the bulge, discs, and halo. The two models are more clearly
distinguished in this plot, primarily because of the larger value of
$R_0$ chosen for the convenient model, which results in a higher value
for $v_0$, because of the strong correlation between the two.

\subsection{Effect of changing the disc scale-heights} \label{sec:height}
In all of the models discussed thus far (and shown in 
Figures~\ref{fig:Rdisc}--\ref{fig:rc} and Tables~\ref{tab:param} \& 
\ref{tab:corr}) the scale-heights of the discs have been held constant
at $300\pc$ and $900\pc$ for the thin and thick discs respectively. 
These values were chosen because they are the best-fitting 
scale-heights given by \cite{Juea08_short}, and were held constant because we 
do not expect the kinematic data described in Section~\ref{sec:data} to 
strongly constrain the vertical density profile. It is, however, 
important that we check this assumption by considering different scale-heights 
for the disc. Therefore we now consider models which have a thin disc 
scale-height $h_\mathrm{d,thin}$ of $250,300$ or $350\pc$ and a thick disc 
scale-height $h_\mathrm{d,thick}$ of 
$750,900$ or $1050\pc$, in all nine possible combinations.

One change to our prior must be considered because of the effect of 
changing the scale-height of the disc. The local density normalisation, 
$f_{\mathrm{d},\odot} =\rho_{\mathrm{thick}}(R_\odot,z_\odot)/ 
\rho_{\mathrm{thin}}(R_\odot,z_\odot)$, is defined at the Sun's position, 
so if the 
scale-heights change, the value of $f_{\mathrm{d},\odot}$ changes without 
the surface densities of the two discs changing. It has been noted
\citep[by, for example,][]{ReRo01} that there is an anti-correlation 
between the values of $f_{\mathrm{d},\odot}$ and $h_\mathrm{d,thick}$ found by 
studies using star counts, which can indeed
be seen in figures 21 \& 24 of \cite{Juea08_short}. We therefore approximate 
that $f_{\mathrm{d},\odot}$ simply changes with $h_\mathrm{d,thick}$, taking the 
values $f_{\mathrm{d},\odot}=0.14,0.12,0.10$ for 
$h_\mathrm{d,thick}=750,900,1050\pc$ respectively, with an uncertainty in 
$f_{\mathrm{d},\odot}$ of $\pm0.012$ in all cases (as previously). These 
values are taken by eye from figure~21 of \cite{Juea08_short}, with a 
correction for the recognised biases in the data. We ignore any 
possible correlation between $f_{\mathrm{d},\odot}$ and $h_\mathrm{d,thin}$ 
 -- an anti-correlation of this type is seen in figures 21 \& 24 of 
\cite{Juea08_short}, but could easily be related to the relationship 
between $f_{\mathrm{d},\odot}$ and $h_\mathrm{d,thick}$, as the values they 
derive for $h_\mathrm{d,thin}$ and $h_\mathrm{d,thick}$ are also correlated.

We give the mean and standard deviation for the parameters (and derived 
quantities) of all these models in the appendix. Here we 
discuss the most significant findings.

Most importantly, the changes in scale heights have very little impact on 
the overall structure of the Galaxy models. The bulge mass and virial mass of 
the Galaxy are virtually unchanged, and the total stellar mass changes by at 
most $\sim5\%$, which is much less than the standard deviation of the value.
The disc scale lengths are also barely changed (changes of at most $\sim 2\%$
in $R_{\mathrm{d},\mathrm{thin}}$ and $\sim 4\%$
in $R_{\mathrm{d},\mathrm{thick}}$). The surface density at the 
Sun is also largely unchanged, with changes of $\sim 1\%$ in 
$K_{z,1.1}$ and the local disc surface
density $\Sigma_{\mathrm{d},\odot}$ changing by $\sim 5\%$. 

The only significant change is in the values of 
$\Sigma_{\mathrm{d},0,\mathrm{thin}}$ and $\Sigma_{\mathrm{d},0,\mathrm{thick}}$, 
which show a transfer of mass from the thick disc to the thin disc
as $h_{\mathrm{d},\mathrm{thin}}$ increases. The mean value of 
$\Sigma_{\mathrm{d},0,\mathrm{thin}}$ rises from $\sim700\msun\pc^{-2}$ for 
models with $h_\mathrm{d,thin}=0.25$ to $\sim800\msun\pc^{-2}$ for those 
with $h_\mathrm{d,thin}=0.35$, while the value of 
$\Sigma_{\mathrm{d},0,\mathrm{thick}}$ declines from $\sim280\msun\pc^{-2}$ to
$\sim210\msun\pc^{-2}$ over the same range. This then affects
the fraction of the total disc mass found in each disc, and 
the value of $f_{\mathrm{d},\odot}$. 

In every case 
the posterior pdf of $f_{\mathrm{d},\odot}$ is almost precisely the same 
as the prior. This indicates that the prior on 
$f_{\mathrm{d},\odot}$ is, in all cases,
the dominant factor determining the how the Galaxy model's total disc mass 
is divided between the two discs.

We can therefore see that holding the disc scale-heights at fixed values 
to produce the statistics in Table~\ref{tab:param} 
does not significantly alter the quoted standard deviations, 
except for parameters related to the division of the disc material into 
the thick and thin discs. Here the true uncertainies are 
significantly larger than the quoted values 
because the uncertainty in scale-heights
must be taken into account.


\section{Comparison to other studies} \label{sec:compare}

The thin disc scale length of our best-fitting model is at the larger
end of the~\cite{Juea08_short} range, which we used as a prior. It is
therefore also larger than the values in the range $2$ to $2.5\kpc$
found by some other recent studies \citep{Ojea99,Chea99,Siea02}, and
the value used for the old (age $>0.15\Gyr$) disc in the Besan\c{c}on
Galaxy model \citep[$2.53\kpc$,][]{Roea03}. On the other hand it lies
close to the scale length found by \cite*{KeDaFa91}, $3\pm0.5\kpc$,
and towards the lower end of the range found by \cite{LCea02}
$3.3^{+0.5}_{-0.4}\kpc$.  The stellar mass of our best-fitting model
is $6.61\times10^{10}\msun$ (with mean and standard deviation
$6.43\pm0.63\times10^{10})$. This compares to the value
$6.1\pm0.5\times10^{10}\msun$ found as a ``back of the envelope''
estimate by \cite{Flea06}. The local disc surface density 
($\Sigma_{\mathrm{d},\odot}=63.9\msun\pc^{-2}$ best-fitting; 
$62.0\pm7.6\msun\pc^{-2}$ mean $\pm$ standard deviation) is somewhat larger 
than suggested by studies which count identified matter in the Solar 
neighbourhood, such as that of \cite{KuGi89} who found 
$\Sigma_{\mathrm{d},\odot}=48\pm8\msun\pc^{-2}$, or \cite{Flea06} who found 
$\Sigma_{\mathrm{d},\odot}\sim49\msun\pc^{-2}$.

Our best-fitting bulge mass is close to the centre of the prior
distribution we take from \cite{BiGe02}. This is significantly larger
than most of the values found by DB98, and comparable to the bulge masses
found by \cite{WiPyDu08}. This is somewhat smaller than the masses
determined by many purely photometric studies
\citep[e.g.][]{Dwea95,PiRo04,LaZyMe02}, but, as noted in
Section~\ref{sec:bulge}, there is an important dependence on
assumptions made about the disc. We note that the stellar mass within the
inner $3\kpc$ in our best-fitting model, $\sim2.4\times10^{10}\msun$,
is very close to the bulge mass found by \cite{PiRo04} assuming a
central hole in the disc: $2.4\pm0.6\times10^{10}\msun$.

The baryonic Tully-Fisher relation describes the relationship between 
total baryonic mass $M_{\mathrm{baryon}}$ and the circular speed $V_\mathrm{c}$ 
at some radius, observed in external galaxies. \cite{McGea10} found 
$\log_{10}M_{\mathrm{baryon}} = 4.0\log_{10}V_\mathrm{c}+1.65$ for disc galaxies, with a 
scatter that is entirely consistent with observational uncertainty,
where $1.1\times V_\mathrm{c}$ was the actual observed circular speed. Our 
models have $\log_{10}M_{\mathrm{baryon}}-(4.0\log_{10}V_\mathrm{c}+1.65) =-0.19\pm0.05$, 
suggesting that the Milky Way 
has a significantly higher rotational speed (or, equivalently, lower 
baryonic mass) than the Tully-Fisher
relation predicts. A similar offset from the baryonic Tully-Fisher 
relationship was found for the Milky Way by \cite{Flea06}.  

In Section~\ref{sec:larger}, we noted the difficulty of finding
rigorous constraints on Galactic structure at large radii, and
described some of problems with commonly cited constraints. It is
still instructive to compare our models to the results of these
previous studies.  The escape speed at the Sun of our best-fitting
model is $622\kms$, with mean and standard deviation over all models
of $606\kms$ and $26\kms$ respectively. This compares to the quoted
$90\%$ confidence interval $498$ to $608\kms$ of \cite{Smea07_short}.  The
mass inside $60\kpc$ is $6.2\times10^{11}\msun$ (best-fitting), with
mean and standard deviation $5.9\pm0.5 \times10^{11}\msun$. This is
rather larger than the \cite{Xuea08_short} value of
$4.0\pm0.7\times10^{11}\msun$. \citeauthor{Xuea08_short} used cosmological
simulations to predict the velocity anisotropy of the BHB population
they were studying -- this leads to predictions of rather high radial
anisotropy, which drives the mass estimate towards lower values than
more isotropic or tangentially anisotropic velocity distributions
would suggest.  DB98 adopted the constraint on $M_{100}$, the mass inside
$100\kpc$, $M_{100}=7\pm2.5\times10^{11}\msun$, based on then-available data.
Our best-fitting model ($M_{100}=9.0\times10^{11}\msun$) and the ensemble
of models as a whole (mean $\pm$ standard deviation:
$8.4\pm0.9\times10^{11}$) fit well within this range. The virial mass
of our best-fitting model $M_{v}=1.40\times10^{12}\msun$ (mean $\pm$
standard deviation: $1.26\pm0.24\times10^{12}\msun$) lies well within
the ranges quoted by \cite{WiEv99} or \cite{Baea06}, and slightly
below the lower quartile of the pdf given by \cite{LiWh08}, but well
above their $95\%$ confidence limit.

\section{Conclusions}
We have presented a simple Bayesian method for applying
photometrically and kinematically derived constraints, and theoretical
understanding of galaxy structure, to parametrized
axisymmetric models of the Galaxy in order to investigate the
distribution of mass in its various components.  We have applied this
method to models with an axisymmetric bulge, exponential discs and an
NFW halo. 

The method we have described is sufficiently general that it
could be applied to any sensible parametrized axisymmetric mass model,
with a wide range of kinematic, photometric, theoretical 
or other constraints. The 
specific constraints we apply are summarised in Table~\ref{tab:constr}.
We have shown that these constraints still allow a wide range of 
Galaxy models, and have found a best-fitting model, as well as a model 
that is best-fitting after key parameters have been fixed at 
convenient values. These models will provide a suitable starting point for 
producing fully dynamical Galaxy models. 

We have also shown that the main 
features of our models are unchanged when we consider different disc 
scale-heights, except to the extent that they alter our prior probability 
distribution on the relative contributions of the thin and 
thick discs, which alters their relative contributions in the models. 
It should therefore be noted that the (already weak) constraints we have
on the \emph{relative} contributions of the two discs are even weaker when the 
uncertainty in disc scale-heights is taken into account. This constraint on the 
ratio of thick to thin disc contributions is almost entirely that from our
\cite{Juea08_short} prior, and a different choice of prior could result in a 
significantly different ratio.

The kinematic data we consider here does not help us to constrain 
the vertical density profile of the Galactic discs, but it is clear 
that kinematic data can be used in combination with star counts to provide 
greater insight into the Galactic potential above the plane 
\citep[e.g.][]{Bu10}, which could then be used to improve these models. 

Applying the need for consistancy in our model 
allows us to find tighter constraints on individual parameters than 
those we begin with. This is particularly noticable in the constraints 
our posterior pdf place on the thin disc scale-length 
($R_{\mathrm{d},\mathrm{thin}}$), the Solar radius ($R_0$), and the 
circular velocity at the Solar radius ($v_0$). 
We find that the Galaxy's dark-matter halo concentration, $c_{v^\prime}$, 
is larger than the average value predicted by simulations. 
The Galaxy's halo is less massive than the expected value from 
cosmological simulations, given the Galaxy's stellar mass (or, 
equivalently, the stellar mass is higher than would be expected). In 
contrast, the stellar mass of the Milky Way is lower than the baryonic
Tully-Fisher relation would suggest given its circular velocity -- 
the discrepency between the two expectations for the stellar mass 
is related to the high concentration of the halo.
In addition to the uncertainty on individual parameters, we are able to find 
the correlations between different parameters demanded by our constraints. 

The results described in this paper are all dependent on the choice of
parametrized model and applied constraints. In particular we have not 
attempted to take account of the effect on the CDM halo of baryonic 
processes, or to consider a CDM halo which is not spherically symmetric. 
The systematic uncertainties on the quantities we describe 
are almost undoubtedly larger than the quoted statistical
uncertainties.

 \section*{Acknowledgments}
 I am very grateful to James Binney for insightful comments, and a careful
 reading of a draft of this paper. I thank other members of the 
 Oxford dynamics group, both past and present, for valuable discussions. 
 This work is supported by a grant from the
 Science and Technology Facilities Council.

\bibliographystyle{mn2e} \bibliography{new_refs}

\appendix
\section{Models with different disc scale heights}
For completeness we now provide a table of the mean and standard deviation of 
the parameters of our models in cases where we consider disc scale-heights 
which differ from our default values. These results are discussed in 
Section~\ref{sec:height}.

\begin{table*}
  \begin{tabular}{ccccccccccccc} 
    & $h_{\mathrm{d},\mathrm{thin}}$ & $h_{\mathrm{d},\mathrm{thick}}$ & $\Sigma_{\mathrm{d},0,\mathrm{thin}}$ & $R_{\mathrm{d},\mathrm{thin}}$ & $\Sigma_{\mathrm{d},0,\mathrm{thick}}$ & $R_{\mathrm{d},\mathrm{thick}} $ & 
    $\rho_{\mathrm{b},0}$ & $\rho_{\mathrm{h},0}$ & $r_\mathrm{h} $ & $R_0 $  & & \\ \hline
Mean & 0.25 & 0.75 & 670 & 3.03 & 280 & 3.20 & 95.4 & 0.013 & 17.4 & 8.28 & & \\
Std. Dev. & - & - & 123 & 0.24 & 125 & 0.56 & 7.0 & 0.006 & 3.9 & 0.16 & & \\ \hline
Mean & 0.25 & 0.90 & 695 & 3.00 & 278 & 3.24 & 95.3 & 0.012 & 17.8 & 8.29 & & \\
Std. Dev. & - & - & 119 & 0.22 & 120 & 0.53 & 6.9 & 0.005 & 3.9 & 0.16 & & \\ \hline
Mean & 0.25 & 1.05 & 713 & 3.01 & 292 & 3.22 & 95.8 & 0.011 & 18.8 & 8.28 & & \\
Std. Dev. & - & - & 124 & 0.23 & 137 & 0.57 & 6.8 & 0.005 & 5.1 & 0.16 & & \\ \hline
Mean & 0.30 & 0.75 & 748 & 2.97 & 232 & 3.31 & 95.3 & 0.011 & 18.3 & 8.28 & & \\
Std. Dev. & - & - & 118 & 0.21 & 113 & 0.60 & 6.8 & 0.005 & 4.2 & 0.16 & & \\ \hline
Mean & 0.30 & 0.90 & $741$ & $3.00$ & $238$ & $3.29$ & $95.5$ & $0.012$ 
& $18.0$ & $8.29$  & & \\
Std. Dev. & - & - &123 & 0.22 & 110 & 0.56 & 6.9 & 0.006 & 4.3 & 0.16 & & \\ \hline 
Mean & 0.30 & 1.05 & 757 & 2.99 & 236 & 3.28 & 95.6 & 0.012 & 18.1 & 8.29 & & \\
Std. Dev. & - & - & 116 & 0.21 & 103 & 0.55 & 6.8 & 0.005 & 4.0 & 0.16 & & \\ \hline
Mean & 0.35 & 0.75 & 777 & 2.97 & 194 & 3.33 & 95.5 & 0.013 & 17.9 & 8.28 & & \\
Std. Dev. & - & - & 127 & 0.21 & 81 & 0.54 & 6.9 & 0.007 & 4.6 & 0.16 & & \\ \hline
Mean & 0.35 & 0.90 & 784 & 2.98 & 208 & 3.30 & 95.6 & 0.012 & 18.2 & 8.29 & & \\
Std. Dev. & - & - & 131 & 0.21 & 87 & 0.55 & 6.9 & 0.006 & 4.4 & 0.16 & & \\ \hline
Mean & 0.35 & 1.05 & 790 & 3.00 & 214 & 3.28 & 95.6 & 0.012 & 18.5 & 8.29 & & \\
Std. Dev. & - & - & 122 & 0.21 & 98 & 0.57 & 6.8 & 0.006 & 5.0 & 0.16 & & \\ \hline
    & $h_{\mathrm{d},\mathrm{thin}}$ & $h_{\mathrm{d},\mathrm{thick}}$ & $v_0$ & $M_\mathrm{b}$ &$M_\ast$  & $M_v $ & $K_{z,1.1}$& $\Sigma_{\mathrm{d},\odot}$  & $\rho_{\ast,\odot}$ & $\rho_{\mathrm{h},\odot}$ & $f_{\mathrm{d},\odot}$ &  $g_{\ast,\odot}/g_{\mathrm{h},\odot}$ \\ \hline
Mean & 0.25 & 0.75 & 239.0 & 8.96 & 62.7 & 1240 & 76.4 & 60.3 & 0.097 & 0.0107 & 0.140 & 1.24 \\
Std. Dev & - & - & 4.7 & 0.65 & 6.2 & 220 & 5.3 & 7.5 & 0.012 & 0.0010 & 0.012 & 0.28 \\ \hline
Mean & 0.25 & 0.90 & 239.3 & 8.94 & 64.0 & 1250 & 76.1 & 61.7 & 0.096 & 0.0106 & 0.120 & 1.27\\
Std. Dev & - & - & 4.6 & 0.64 & 6.2 & 220 & 5.2 & 7.5 & 0.012 & 0.0010 & 0.012 & 0.28 \\ \hline
Mean & 0.25 & 1.05 & 238.9 & 8.99 & 65.7 & 1280 & 76.6 & 63.7 & 0.099 & 0.0104 & 0.101 & 1.36 \\
Std. Dev & - & - & 5.0 & 0.64 & 6.4 & 260 & 5.2 & 7.7 & 0.012 & 0.0011 & 0.012 & 0.33 \\ \hline
Mean & 0.30 & 0.75 & 238.7 & 8.94 & 63.7 & 1280 & 77.1 & 61.4 & 0.086 & 0.0106 & 0.140 & 1.30 \\
Std. Dev & - & - & 4.8 & 0.64 & 5.9 & 230 & 5.1 & 7.1 & 0.010 & 0.0010 & 0.012 & 0.29 \\ \hline
Mean & 0.30 & 0.90 &  $239.2$ & $8.96$ & $64.3$ & $1260$ & 76.5 & 62.0 & 0.085 & 0.0106 & 0.120 & 1.29 \\
    Std. Dev.  & - & - & 4.8 & 0.65 & 6.3 & 240 & 5.3 & 7.6 & 0.010 & 0.0010 & 0.012 & 0.30 \\ \hline
Mean & 0.30 & 1.05 & 239.3 & 8.97 & 65.0 & 1260 & 76.1 & 62.7 & 0.085 & 0.0105 & 0.100 & 1.31\\
Std. Dev & - & - & 4.8 & 0.64 & 6.5 & 220 & 5.2 & 7.7 & 0.010 & 0.0010 & 0.012 & 0.30 \\ \hline
Mean & 0.35 & 0.75 & 239.0 & 8.96 & 63.5 & 1260 & 76.5 & 61.1 & 0.077 & 0.0106 & 0.140 & 1.28 \\
Std. Dev & - & - & 4.9 & 0.65 & 6.3 & 240 & 5.3 & 7.6 & 0.010 & 0.0011 & 0.012 & 0.31 \\ \hline
Mean & 0.35 & 0.90 & 239.2 & 8.97 & 64.9 & 1260 & 76.5 & 62.5 & 0.076 & 0.0105 & 0.120 & 1.32\\
Std. Dev & - & - & 4.9 & 0.65 & 6.6 & 240 & 5.4 & 7.8 & 0.010 & 0.0011 & 0.012 & 0.31 \\ \hline
Mean & 0.35 & 1.05 & 239.3 & 8.97 & 65.7 & 1270 & 76.4 & 63.5 & 0.077 & 0.0104 & 0.100 & 1.35 \\
Std. Dev & - & - & 4.8 & 0.64 & 6.8 & 260 & 5.4 & 8.1 & 0.010 & 0.0011 & 0.012 & 0.34 \\ \hline
  \end{tabular}
  \caption{
    Mean and standard deviation of the parameters (upper) and derived 
    properties (lower) 
    of our models, for various values of the thin and thick disc scale-heights
    ($h_{\mathrm{d},\mathrm{thin}}$ \& $h_{\mathrm{d},\mathrm{thick}}$ respectively). 
    This is very similar to Table~\ref{tab:param}. The results with our 
    standard scale-heights ($h_{\mathrm{d},\mathrm{thin}}=0.3\kpc$, 
    $h_{\mathrm{d},\mathrm{thick}}=0.9\kpc$) are included in this table as well 
    as in Table~\ref{tab:param}. Again, distances 
    are quoted in units of $\kpc$, velocities in $\kms$, 
    masses in $10^9\msun$, surface densities in $\msun\pc^{-2}$,
    densities in $\msun\pc^{-3}$, and $K_{z,1.1}$ in units of 
    $(2\pi G)\times\msun\pc^{-2}$.  
  }
\end{table*}

\end{document}